\documentclass[10pt,conference,epsfig]{IEEEtran}
\usepackage[dvips]{graphicx}
\usepackage{graphicx}
\usepackage{amssymb}
\usepackage{cite}
\usepackage{subfigure}
\usepackage{algorithm}
\usepackage{algpseudocode}
\usepackage{amsmath}

\usepackage{subeqnarray}
\usepackage{cases}
\usepackage[Symbol]{upgreek}
\makeatletter

\newcommand{\Rmnum}[1]{\expandafter\@slowromancap\romannumeral #1@}
\makeatother

\begin{document}
\IEEEoverridecommandlockouts
\title{Transmit Design for MIMO Wiretap Channel with a Malicious Jammer
\thanks{This work was supported in part by the Sichuan Province Science and technology program under Grant 2013GZ0029, and by the Important National Science and Technology Specific Projects of China under Grant 2014ZX03004003.}}
\author{\authorblockN{Duo Zhang, Weidong Mei, Lingxiang Li, and Zhi Chen}
\authorblockA{National Key Laboratory of Science and Technology on Communications\\
University of Electronic Science and Technology of China, Chengdu 611731, China\\
Emails: zhangduo1st@gmail.com; mwduestc@gmail.com; walk22talk@gmail.com; chenzhi@uestc.edu.cn;}}
\maketitle

\begin{abstract}
In this paper, we consider the transmit design for multi-input multi-output (MIMO) wiretap channel including a malicious jammer. We first transform the system model into the traditional three-node wiretap channel by whitening the interference at the legitimate user. Additionally, the eavesdropper channel state information (ECSI) may be fully or statistically known, even unknown to the transmitter. Hence, some strategies are proposed in terms of different levels of ECSI available to the transmitter in our paper. For the case of unknown ECSI, a target rate for the legitimate user is first specified. And then an inverse water-filling algorithm is put forward to find the optimal power allocation for each information symbol, with a stepwise search being used to adjust the spatial dimension allocated to artificial noise (AN) such that the target rate is achievable. As for the case of statistical ECSI, several simulated channels are randomly generated according to the distribution of ECSI. We show that the ergodic secrecy capacity can be approximated as the average secrecy capacity of these simulated channels. Through maximizing this average secrecy capacity, we can obtain a feasible power and spatial dimension allocation scheme by using one dimension search. Finally, numerical results reveal the effectiveness and computational efficiency of our algorithms.
\end{abstract}

\section{Introduction}
Due to the broadcast nature of wireless medium, physical layer security approach is playing an increasingly important role in wireless communication recently. It can achieve significant security performance without using secret keys whose distribution and management may lead to security holes in wireless systems. Wyner's work in \cite{wyner1975wire} has established the fundamentals of physical-layer secrecy since 1970's, in which the positive secrecy rate is proved to be achievable in a degraded wiretap channel; refer to \cite{hong2013signal} for a detailed overview of physical-layer secrecy. The past several decades have witnessed a rapid development of research on physical-layer secrecy, especially in the multi-input multi-output (MIMO) channels.  For ease of exposition, we name the transmitter, the legitimate receiver and the eavesdropper \emph{Alice}, \emph{Bob} and \emph{Eve}, respectively.

Different tactics against eavesdropper have been developed with various levels of eavesdropper channel state information (ECSI) available to Alice. For the case of full ECSI, Khisti and Wornell extended Wyner's work to the multi-input, single-output, multi-eavesdropper (MISOME) and multi-input, multi-output, multi-eavesdropper (MIMOME) wiretap channel, respectively \cite{khisti2010secure1, khisti2010secure2}. Furthermore, some effective transmit designs are proposed to maximize the secrecy rate by adopting alternating optimization (AO) algorithm and generalized singular value decomposition (GSVD) algorithm in \cite{li2013transmit} and \cite{fakoorian2012optimal}, respectively. For the case of unknown ECSI, Goel and Negi first proposed the technique of artificial noise (AN) to degrade Eve's channel \cite{goel2008guaranteeing}. While for the case of statistical ECSI, it is more intricate due to being less mature in comparison with that on full CSI and unknown CSI. Several works have focused on the statistical CSI recently. J. Li and A. P. Petropulu have proved the optimality of beamforming in the \textbf{MISOME} system where Eve's channel may have a nonzero mean or nontrivial covariance matrix \cite{li2011onergodic, li2011ergodic}. In \cite{zhou2010secure}, AN is taken into account in \textbf{MISOME} system, and the optimal power allocation strategy is determined by attaining the lower bound of the ergodic secrecy capacity. Moreover, an iterative algorithm is firstly proposed to deal with the \textbf{MIMOME} system with statistical ECSI in \cite{zhu2013artificial}. On the other hand, the impact of malicious jammer on the secrecy performance is another problem of long-standing interest. Seeking for the effective tactics against malicious jammer turns out urgent and critical in some special networks, such as, military networks. A prevalent approach is to model the transmitter and malicious jammer as players in a game-theoretic formulation \cite{mukherjee2013jamming}. Nevertheless, these game-theoretic-based works seldom concentrate on the optimal power allocation strategy for both parties. Actually, in order to degrade Eve's channel and guarantee Bob's reception quality simultaneously, the power allocation strategies turn out to be a dilemma. To the best of our knowledge, few studies work on the scenario for its challenges.

In this paper, we consider the transmit design for four-node multiple-antenna wiretap channel including a malicious jammer. We firstly show that the four-node wiretap channel can be transformed into the traditional three-node wiretap channel by whitening the interference at Bob. Based on the channel's transformation, different transmit schemes are proposed with various levels of ECSI available at Alice. Specifically, for the case of unknown ECSI, a novel optimization algorithm is put forward. A target rate for Bob is first set, and an inverse water-filling algorithm is derived to determine the optimal power allocation scheme for Alice. Meanwhile, a stepwise search is then used to find the optimal spatial dimension allocated to AN such that the target rate is achievable. For the case of statistical ECSI, in view of the high computational complexity of the scheme proposed in \cite{zhu2013artificial}, we simulate some random channel according to the distribution of ECSI. Then the ergodic secrecy capacity can be approximated as the average secrecy capacity of these simulated channels. To maximize the average secrecy capacity, we employ one dimension search to determine the optimal power allocation and optimal spatial dimension allocation between the information signal and AN. The numerical results verify the efficacy of our proposed schemes.

\emph{Notation}: Bold symbols in capital letter and small letter denote matrices and vectors, respectively. ${{(\bullet)}^{\text{H}}}$, $\text{Tr}(\bullet )$, ${{(\bullet )}^{-\text{1}}}$ and $\left| \bullet  \right|$ represent conjugate transpose, trace, inversion and determinant of a matrix. ${\mathbb{Z}}_{+}$ is the set of nonnegative integers. $\mathbf{I}$ denotes an identity matrix and ${{[x]}^{+}}=\text{max}(0,x)$. $\mathbf{x}\sim \mathcal{CN}(\mathbf{\mu },\mathbf{\Omega })$ denotes that \textbf{x} is a complex circular Gaussian random vector with mean $\mathbf{\mu}$ and covariance $\mathbf{\Omega}$. $\mathbf{A}\succeq \mathbf{0}$ implies that \textbf{A} is a Hermitian positive semidefinite (PSD) matrix while $\mathbf{x}\succeq \mathbf{0}$ implies that each component of \textbf{x} is nonnegative.

\section{System Model}
\begin{figure}[!t]
\centering
\includegraphics[scale=0.6]{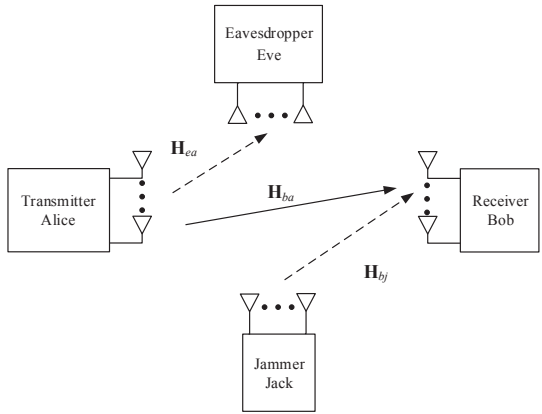}
\DeclareGraphicsExtensions. \caption{MIMO wiretap channel with a malicious jammer}\label{fig 1}
\end{figure}
Consider a four-node multiple-antenna wiretap channel including a malicious jammer, which is illustrated in Fig.\,\ref{fig 1}. All terminals are equipped with multiple antennas. We will call the malicious jammer \emph{Jack} for short. Suppose that Alice, Bob, Eve and Jack are equipped with \emph{N}$_A$, \emph{N}$_B$, \emph{N}$_E$ and \emph{N}$_J$ antennas, respectively. It is assumed that Jack only transmits AN to degrade Bob's reception and no information will be shared between Jack and Eve. The received signals at Bob and Eve can be modeled as
\begin{align}
&{{\mathbf{y}}_{b}}={{\mathbf{H}}_{ba}}{{\mathbf{x}}_{a}}+{{\mathbf{H}}_{bj}}\mathbf{j}+{{\mathbf{n}}_{b}}\label{yb}\\
&{{\mathbf{y}}_{e}}={{\mathbf{H}}_{ea}}{{\mathbf{x}}_{a}}+{{\mathbf{n}}_{e}}\label{ye}
\end{align}
respectively, where ${{\mathbf{H}}_{ba}}\in {{\mathbb{C}}^{{{N}_{B}}\times {{N}_{A}}}}$ and ${{\mathbf{H}}_{bj}}\in {{\mathbb{C}}^{{{N}_{B}}\times {{N}_{J}}}}$ represent the MIMO channel from Alice and Jack to Bob, respectively,and ${{\mathbf{H}}_{ea}}\in {{\mathbb{C}}^{{{N}_{E}}\times {{N}_{A}}}}$ represents the MIMO channel from Alice to Eve. ${{\mathbf{x}}_{a}}\in {{\mathbb{C}}^{{{N}_{A}}}}$ is the signal transmitted by Alice. $\mathbf{j}\in {{\mathbb{C}}^{{{N}_{J}}}}$ is the AN generated by Jack to jam Bob. Without loss of generality, the additive noise, ${{\mathbf{n}}_{b}}\in {{\mathbb{C}}^{{{N}_{B}}}}$ and ${{\mathbf{n}}_{e}}\in {{\mathbb{C}}^{{{N}_{E}}}}$, are supposed as i.i.d. complex Gaussian noise with zero mean and unit variance. Since ${{\mathbf{H}}_{bj}}$ is fixed over a sufficiently long period of time, Bob can detect it and then feed it back to Alice. It is reasonable to assume that ${{\mathbf{H}}_{ba}}$ and ${{\mathbf{H}}_{bj}}$ are both known to Alice, but ${{\mathbf{H}}_{ea}}$ may be only fully or statistically known, even unknown to Alice.

Suppose that ${{\mathbf{H}}_{bj}}$ is inaccessible to Jack, so it is an optimal strategy for Jack to transmit omni-directional AN \cite{bayesteh2004effect}, i.e.,
\begin{equation}\label{Qj}
  {{\mathbf{Q}}_{j}}=\frac{{{P}_{j}}}{{{N}_{j}}}\mathbf{I}
\end{equation}
where ${{\mathbf{Q}}_{j}}$ is the covariance matrix of $\mathbf{j}$, $P_j$ is the total transmission power of Jack.

In this paper, we focus on the optimization of the input covariance matrix design at Alice. The four-node-based secrecy rate maximization (SRM) design can be formulated as
\begin{equation}\label{Rs1}
\begin{split}
  & R_{s}^{*}=\max \ {{\left[ {{R}_{b}}-{{R}_{e}} \right]}^{+}} \\
 & s.t.\quad \text{Tr(}{{\mathbf{Q}}_{a}})\le P
 \end{split}
\end{equation}
where $R_b$ and $R_e$ denote the achievable transmission rate between Alice and Bob/Eve, i.e.,
\begin{align}
&{{R}_{b}}=\ln \left| \mathbf{I}+{{(\mathbf{I}+{{\mathbf{H}}_{bj}}{{\mathbf{Q}}_{j}}{{\mathbf{H}}_{bj}}^{H})}^{-1}}{{\mathbf{H}}_{ba}}{{\mathbf{Q}}_{a}}{{\mathbf{H}}_{ba}}^{H} \right|\\
&{{R}_{e}}=\ln \left| \mathbf{I}+{{\mathbf{H}}_{\text{e}a}}{{\mathbf{Q}}_{a}}{{\mathbf{H}}_{ea}}^{H} \right|
\end{align}
in which ${{\bf{Q}}_a} \buildrel \Delta \over = E({{\bf{x}}_a}{\bf{x}}_a^H)$ is the covariance matrix of $\bf{x}_a$.

Because the problem above is nonconvex, it's generally difficult to obtain the optimal solution. Note that when ECSI is unknown, this problem gets further intractable to deal with. To tackle it, we will simplify this problem by utilizing the interference whitening in the next section.

\section{Equivalent System Model}
We define
\begin{equation}\label{CC}
  {{(\mathbf{I}+{{\mathbf{H}}_{bj}}{{\mathbf{Q}}_{j}}{{\mathbf{H}}_{bj}}^{H})}^{-1}}=\mathbf{C}{{\mathbf{C}}^{H}}
\end{equation}
for its positive semi-definition. According to the Sylvester's determinant theorem, that is, $\det ({{\bf{I}}_m} + {\bf{AB}}) = \det ({{\bf{I}}_n} + {\bf{BA}})$, for any ${\mathbf{A}}\in {{\mathbb{C}}^{m \times n}}$ and ${\mathbf{B}}\in {{\mathbb{C}}^{n \times m}}$, $R_b$ can thus be simplified as
\begin{equation}\label{Rbsim}
  {{R}_{b}}=\ln \left| \mathbf{I}+{{{\mathbf{\tilde{H}}}}_{ba}}{{\mathbf{Q}}_{a}}\mathbf{\tilde{H}}_{ba}^{H} \right|
\end{equation}
where ${{\mathbf{\tilde{H}}}_{ba}}\triangleq {{\mathbf{C}}^{H}}{{\mathbf{H}}_{ba}}$ is defined as the equivalent channel matrix.

\begin{figure}[!t]
\centering
\includegraphics[scale=0.6]{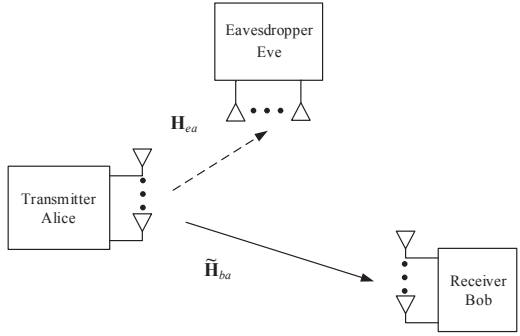}
\DeclareGraphicsExtensions.
\caption{The equivalent MIMO wiretap channel} \label{fig 2}
\end{figure}

In doing this way, the four-node wiretap channel with a jammer has been transformed into the conventional three-node wiretap channel, which is depicted in Fig.\,\ref{fig 2}. The equivalent received signals at Bob and Eve are given in (\ref{req}).
\begin{equation}\label{req}
\begin{split}
& {{{\mathbf{\tilde{y}}}}_{b}}={{{\mathbf{\tilde{H}}}}_{ba}}{{\mathbf{x}}_{a}}+{{{\mathbf{\tilde{n}}}}_{b}} \\
& {{\mathbf{y}}_{e}}={{\mathbf{H}}_{ea}}{{\mathbf{x}}_{a}}+{{\mathbf{n}}_{e}}
\end{split}
\end{equation}
In (\ref{req}), ${{\mathbf{\tilde{y}}}_{b}}\triangleq {{\mathbf{C}}^{H}}{{\mathbf{y}}_{b}}$ and ${{\mathbf{\tilde{n}}}_{b}}\triangleq {{\mathbf{C}}^{H}}({{\mathbf{H}}_{bj}}\mathbf{j}+{{\mathbf{n}}_{b}})$ are defined as the equivalent received signal and white noise at Bob, ${{\mathbf{\tilde{n}}}_{b}}\sim \mathcal{CN}(\mathbf{0},\mathbf{I})$.

\section{Transmit Design with Different Levels of ECSI}
\subsection{Unknown ECSI at Alice}
In many applications, it is impractical to assume that full ECSI is available to Alice. When the ECSI is unknown to Alice, it is infeasible to solve the problem in (\ref{Rs1}) directly. In \cite{mukherjee2011robust}, it has been indicated that maximizing the AN can improve the secrecy performance as long as a desired SINR is guaranteed to achieve. However, the scheme proposed in \cite{mukherjee2011robust} merely aims at single-flow data transmission, that is, only one spatial dimension is exploited for information symbols while the others are exploited for AN. In fact, we are able to degrade Eve's reception subspace with only $N_E$ spatial dimensions being allocated to AN. Therefore, in this subsection, we put forward another algorithm which makes effective use of spatial dimension to further improve the secrecy performance.
To be specific, we first specify a target data rate $R$ for Bob. Then $N_e$ dimensions are allocated to AN which is generated to be orthogonal with the information symbols at Bob. Next, a proper scheme is designed to achieve the target data rate with the smallest possible power allocated to the information symbols. Particularly, it should be mentioned that some target rate may be unreachable in some cases, in such cases, the communication is assumed to be in outage. Apparently, the more power allocated to AN, the more degradation Eve will encounter. But one can notice that the secrecy rate will also decrease resulted from less power allocated to information symbols. Hence, the power distribution ratio between information and AN, denoted by $\rho$, should be determined prudently.

With AN being transmitted, the transmit signal ${{\mathbf{x}}_{a}}$ can be decomposed into the following form,
\begin{equation}\label{Xa}
{{\mathbf{x}}_{a}}={{\mathbf{W}}_{u}}\mathbf{u}+{{\mathbf{W}}_{z}}\mathbf{z}
\end{equation}
where $\mathbf{u}\in {{\mathbb{C}}^{{{d}_{1}}}}$ is information-bearing signal intended for Bob, $\mathbf{z}\in {{\mathbb{C}}^{{{d}_{2}}}}$ is the AN created by Alice. The elements of $\mathbf{z}$ are chosen to be i.i.d. complex Gaussian random variables. Assume that $\mathbf{u}\sim \mathcal{CN}(0,{{\mathbf{Q}}_{u}})$ and $\mathbf{z}\sim \mathcal{CN}(0,{{\mathbf{Q}}_{z}})$, $\mathbf{u}$ and $\mathbf{z}$ are assumed to be independent. ${{\mathbf{W}}_{u}}\in {{\mathbb{C}}^{{{N}_{A}}\times {{d}_{1}}}}$ and ${{\mathbf{W}}_{z}}\in {{\mathbb{C}}^{{{N}_{A}}\times {{d}_{2}}}}$ represent the precoding matrix and the AN shaping matrix respectively, in which $d_1$ and $d_2$ represent the spatial dimension allocated for $\mathbf{u}$ and $\mathbf{z}$. To avoid the AN interference at Bob, $({{\mathbf{W}}_{u}},{{\mathbf{W}}_{z}})$ should be an orthogonal basis of ${{\mathbb{C}}^{{{N}_{A}}}}$.

We consider the total power budget denoted by $P$, and then we arrive at
\begin{equation}\label{Q}
\begin{split}
 & {{\mathbf{Q}}_{u}}\text{=diag}({{P}_{1}},{{P}_{2}},\cdots ,{{P}_{{{d}_{1}}}}),\text{Tr(}{{\mathbf{Q}}_{u}}\text{)=}\rho P \\
 & {{\mathbf{Q}}_{z}}\text{=}\frac{\text{(1}-\rho )P}{{{d}_{2}}}\mathbf{I},\text{Tr(}{{\mathbf{Q}}_{z}}\text{)=(1}-\rho )P
\end{split}
\end{equation}
where ${P_i}(i = 1,2,...,{d_1})$ denotes the power allocated to $i$th information symbol. Note that the AN power is uniformly distributed to each AN symbol, which has been proved to be the best option for Alice in \cite{liao2011qos}.

Since the ECSI is unknown, Alice is assumed to begin with $N_E$ spatial dimension being allocated to AN to degrade Eve's reception. Meanwhile, more dimensions will be allocated to information symbols if the target rate is too high to achieve. As a result, if $c$ is assumed to be the number of nonzero singular value of ${{\mathbf{\tilde{H}}}_{ba}}$, Alice should choose $\mathbf{W}_u$ as the right singular vectors corresponding to the $\min({N_A} - {N_E},c)$ largest singular values of ${{\mathbf{\tilde{H}}}_{ba}}$. Here, we suppose that Alice is equipped with more antennas than Eve, i.e., $N_A>N_E$. $\mathbf{W}_z$ must be composed of the remaining $N_A-d_1$ right singular vectors of ${{\mathbf{\tilde{H}}}_{ba}}$. Hence, we have
\begin{equation}\label{datastream}
\begin{split}
 & {d_1} =\min({N_A} - {N_E},c) \\
 & {{d}_{2}}={{N}_{A}}-{{d}_{1}}
\end{split}
\end{equation}
Then the real data rate for Bob is given by
\begin{equation}\label{Rbwf}
 {{R}_{b}}=\sum\limits_{i=1}^{{{d}_{1}}}{\ln (1+\sigma _{i}^{2}{{P}_{i}})}
\end{equation}
where $\left\{ {{\sigma }_{i}} \right\}_{i=1}^{{{d}_{1}}}$ are the $d_1$ largest nonzero singular values of ${{\mathbf{\tilde{H}}}_{ba}}$.

To minimize the fraction of the transmit power required to achieve the target rate $R$, a power allocation minimum design can be formulated as
\begin{equation}\label{func_or}
\begin{split}	
  & \underset{{{P}_{i}},i=1,...,{{d}_{1}}}{\mathop{\min }}\,\sum\limits_{i=1}^{{{d}_{1}}}{{{P}_{i}}} \\
 s.t.& \quad\sum\limits_{i=1}^{{{d}_{1}}}{\ln (1+\sigma _{i}^{2}{{P}_{i}})}=R \\
 & \quad\mathbf{p}\triangleq {{[{{P}_{1}},{{P}_{2}},...,{{P}_{{{d}_{1}}}}]}^{T}}\succeq\mathbf{0} \\
 & \quad\sum\limits_{i=1}^{{{d}_{1}}}{{{P}_{i}}}\le P
 \end{split}
 \end{equation}
Recalling the conventional water-filling algorithm, one can find that (\ref{func_or}) operates in an inverse way. Hence, we name our algorithm as an inverse water-filling algorithm.

Let ${{P}_{i}}=\frac{{{e}^{{{x}_{i}}}}-1}{\sigma _{i}^{2}}$, for $i=1,...,d_1$. Problem (\ref{func_or}) can be equivalently expressed as
 \begin{equation}\label{func_trans}
\begin{split}
  & \underset{{{x}_{i}},i=1,...,{{d}_{1}}}{\mathop{\min }}\,\sum\limits_{i=1}^{{{d}_{1}}}{\frac{{{e}^{{{x}_{i}}}}}{\sigma _{i}^{2}}} \\
 s.t.&\quad \sum\limits_{i=1}^{{{d}_{1}}}{{{x}_{i}}}=R \\
 &\quad -\mathbf{x}\triangleq -{{[{{x}_{1}},{{x}_{2}},...,{{x}_{{{d}_{1}}}}]}^{T}}\preceq \mathbf{0} \\
 &\quad \sum\limits_{i=1}^{{{d}_{1}}}{\frac{{{e}^{{{x}_{i}}}}}{\sigma _{i}^{2}}}\le P+\sum\limits_{i=1}^{{{d}_{1}}}{\frac{1}{\sigma _{i}^{2}}}
\end{split}
\end{equation}
It's easy to verify that the power allocation minimum problem above is convex with respect to $\left\{ {{x}_{i}} \right\}_{i=1}^{{{d}_{1}}}$.

However, an intractable problem the power allocation minimum design may incur is that, the target rate could be unreachable under the total power constraint. To tackle it, let us consider a relaxed power allocation minimum problem where the total power constraint is dropped. By doing so, if the solution of the relaxed power allocation minimum problem does not satisfy the total power constraint, then more dimensions will be allocated to the information signal. Particularly, if the target rate is still unreachable when all the power has been allocated to the information signal, we will break off the link directly.

To solve the relaxed power allocation minimum problem, we consider the following Lagrangian of it,
\begin{equation}\label{larg}
L(\mathbf{x},\mathbf{\uplambda },\nu )=\sum\limits_{i=1}^{{{d}_{1}}}{\frac{{{e}^{{{x}_{i}}}}}{\sigma _{i}^{2}}}-{{\mathbf{\uplambda }}^{T}}\mathbf{x}+\nu (\sum\limits_{i=1}^{{{d}_{1}}}{{{x}_{i}}}-R)
\end{equation}
where ${\mathbf{\lambda }}$ and $\nu$ are the Lagrange multipliers associated with the constraint $\mathbf{x}\succeq\mathbf{0}$ and $\sum\limits_{i=1}^{{{d}_{1}}}{{{x}_{i}}}=R$, respectively. By Slater's condition \cite{boyd2009convex}, the strong duality holds true for the \textbf{relaxed} power allocation minimum problem. Then the optimal $\mathbf{x}$ can be obtained from the KKT conditions as below.
\begin{align}
  \frac{\partial \sum\limits_{i=1}^{{{d}_{1}}}{\frac{{{e}^{{{x}_{i}}}}}{\sigma _{i}^{2}}}}{\partial {{x}_{i}}}-\frac{\partial {{\mathbf{\lambda }}^{T}}\mathbf{x}}{\partial {{x}_{i}}}+\frac{\partial v(\sum\limits_{i=1}^{{{d}_{1}}}{{{x}_{i}}}-R)}{\partial {{x}_{i}}}=0& \label{kkt1}\\
 \sum\limits_{i=1}^{{{d}_{1}}}{{{x}_{i}}}-R=0& \label{kkt2}\\
{{\lambda }_{i}}\ge 0 & \label{kkt3}\\
 -{{\lambda }_{i}}{{x}_{i}}=0&\label{kkt4}
\end{align}
which yields
\begin{equation}\label{kkt1s}
\frac{{{e}^{{{x}_{i}}}}}{\sigma _{i}^{2}}+v={{\lambda }_{i}}
\end{equation}
 By (\ref{kkt2})-(\ref{kkt1s}), we consider the following two cases to solve the optimal $P_i$.
\begin{itemize}
  \item ${{\lambda }_{i}}>0$ and ${{x}_{i}}=0\Rightarrow {{\lambda }_{i}}=\frac{1}{\sigma _{i}^{2}}+\nu >0\Rightarrow \ln (-v\sigma _{i}^{2})<0$
  \item ${{\lambda }_{i}}=0$ and ${{x}_{i}}\ge 0\Rightarrow {{x}_{i}}=\ln (-v\sigma _{i}^{2})\ge0$
\end{itemize}

It can be extracted from the two above cases that
\begin{equation}\label{optallo}
{{x}_{i}}^{opt}=\max (0,\ln (-v\sigma _{i}^{2}))\Rightarrow P_{i}^{opt}=\max (0,-v-\frac{1}{\sigma _{i}^{2}})
\end{equation}
The optimal $\nu$ can be solved from (\ref{kkt2}) in the similar way as conventional water-filling algorithm. The overall procedures proposed for the unknown ECSI case have been summarized in \textbf{Algorithm} 1.
\begin{algorithm}
  \caption{The optimal strategy for Alice without ECSI}
  \begin{algorithmic}[1]
    \State Initialize the target data rate $R$, the number of Eve's antennas $N_E$, the total power budget $P$.
    \State Compute the singular value decomposition of ${{\mathbf{\tilde{H}}}_{ba}}$.
    \State Compute ${{d}_{1}}=\min({{N}_{A}}-{{N}_{E}},c)$, ${{d}_{2}}={{N}_{A}}-{{d}_{1}}$, then determine $\mathbf{W}_{u}^{{}}$,$\mathbf{W}_{z}^{{}}$ according to the tactics above (\ref{datastream}).
    \State Obtain ${{P}_{i}}^{opt}$ and $\rho$ by ${{P}_{i}}^{opt}=\max (0,-v-\frac{1}{{{\sigma }_{i}^{2}}})$ and $\rho =(\sum\limits_{i=1}^{{{d}_{1}}}{{{P}_{i}}})/P$, respectively.
    \State \textbf{while} $\rho >1$
    \State ${{d}_{1}}={{d}_{1}}+1,{{d}_{2}}={{d}_{2}}-1$, renew $\mathbf{W}_{u}^{{}}$ and $\mathbf{W}_{z}^{{}}$ by the tactics above (\ref{datastream}).
    \State Recompute ${{P}_{i}}^{opt}$ and $\rho $.
    \State \textbf{end while}
    \State \textbf{if} $\rho >1$
    \State Break off the link, and ${{R}_{b}}=0$.
    \State \textbf{else}
    \State Output ${{P}_{i}}^{opt}$,$\rho $.
    \State \textbf{end if}
  \end{algorithmic}
\end{algorithm}

In particular, it should be reminded that for the case of unknown $N_e$, one can start by initializing $d_1=1$. The secrecy rate obtained from \textbf{Algorithm 1} is given by
\begin{equation}
{R_s} = \;\left\{ \begin{array}{l}
R - \ln \left| {{\bf{I}} + {\bf{\Phi }}} \right|{\kern 1pt} \quad {\rm{if}}\;R\;{\rm{is}}\;{\rm{achievable}}\\
0\quad \quad \quad \quad \quad \quad \;\;\;{\rm{otherwise}}
\end{array} \right.
\end{equation}
where $\mathbf{\Phi }={{\left( \mathbf{I}+{{\mathbf{H}}_{ea}}\mathbf{W}{{\mathbf{Q}}_{{z}}}{{\mathbf{W}}^{H}}\mathbf{H}_{ea}^{H} \right)}^{-1}}{{\mathbf{H}}_{\text{e}a}}\mathbf{V}{{\mathbf{Q}}_{u}}{{\mathbf{V}}^{H}}\mathbf{H}_{ea}^{H}$, in which ${{\mathbf{Q}}_{z}}$ and ${{\mathbf{Q}}_{u}}$ are given by (\ref{Q}), and ${{\mathbf{Q}}_{u}}$'s diagonal elements are composed of $\left\{ P_{i}^{opt} \right\}_{i=1}^{{{d}_{1}}}$.

\subsection{Statistical ECSI at Alice}
When only the statistical ECSI is available to Alice, a general approach is to consider the ergodic secrecy rate maximization (ESRM) problem. However, the problem of ERSM is challenging for the barriers of solving the optimization problem. Alternatively, one can choose to concentrate on maximizing the lower bound of ergodic secrecy rate. In \cite{zhu2013artificial}, this problem has been derived by \emph{Yan Zhu} \emph{et.al}. However, the iterative algorithm proposed in \cite{zhu2013artificial} are unpractical and inefficient for its high complexity to compute the lower bound. Instead, in this subsection, we put forward an alternative solution to tackle the case of statistical ECSI.

Consider that Alice has an access to the statistical ECSI, we employed a stochastic simulation method. Specifically, some simulated channel between Alice and Eve can be randomly generated based on the distribution of $\mathbf{H}_{ea}$. The key idea is that, from the characteristics of Gaussian distribution, the simulated channels will assemble in a restricted interval with high probability. Hence, the estimation error will also be restricted in an acceptable scope. And then, the ergodic secrecy capacity can be approximated as the average secrecy capacity of the simulated channels. To this end, a great number of simulated channels are expected to generate to reduce the average error. According to our later numerical results, we can attain a secrecy performance identical to \cite{zhu2013artificial} within five channel trials.

AN is assumed to be transmitted to degrade Eve's reception. We number the simulated channels as ${{{{\mathbf{\hat H}}}_{ea}}(i)}$, $i=1,2,...,N_{sim}$, $N_{sim}$ denotes the total number of simulated channels. Similarly to the tactics for unknown ECSI, the precoding matrix ${{\mathbf{W}}_{u}}$ is assumed to be composed of the right singular vectors corresponding to the $r$ largest singular values of ${{\mathbf{\tilde{H}}}_{ba}}$. The AN shaping matrix ${{\mathbf{W}}_{z}}$ is composed of the remaining $N_A-r$ right singular vectors of ${{\mathbf{\tilde{H}}}_{ba}}$. We suppose the power distribution ratio between information and AN is denoted as $\rho$ yet. Then the secrecy rate for $i$th simulated channel is given by ($i=1,2,...,N_{sim}$)
\begin{equation}\label{Rsim}
{{\hat R}_s(r,\rho,i)} = \ln \left| {{\bf{I}} + {{\bf{H}}_{ba}}{\bf{V}}{{\bf{Q}}_u}{{\bf{V}}^H}{\bf{H}}_{ba}^H} \right| - \ln \left| {{\bf{I}} + {\bf{\hat \Phi }}(i)} \right|
\end{equation}
where
\[{\bf{\hat \Phi }}(i) = {\left( {{\bf{I}} + {{\bf{H}}_{ea}}{\bf{W}}{{\bf{Q}}_z}{{\bf{W}}^H}{\bf{H}}_{ea}^H} \right)^{ - 1}}{{{\bf{\hat H}}}_{ea}}(i){\bf{V}}{{\bf{Q}}_u}{{\bf{V}}^H}{\bf{\hat H}}_{ea}^H(i),\]
and ${{\bf{Q}}_u}$ and ${{\bf{Q}}_z}$ can be expressed by
\begin{equation}
{{\bf{Q}}_u} = \rho P \cdot {\rm{waterfilling}}(\rho ,r),{{\bf{Q}}_z} = \frac{{{\rm{(1}} - \rho )P}}{{{N_A} - r}}{\bf{I}}
\end{equation}
in which the optimal power allocation matrix for the information symbols is obtained via the water-filling algorithm. The global suboptimal $\rho^*$ and $r^*$ are thereby attained by maximizing the average secrecy capacity of the simulated channels. The overall procedures for the statistical ECSI case have been summarized in \textbf{Algorithm 2}.

The search for $\rho^*$ can be implemented by adopting the one dimension search algorithms, such as, golden section search and bisection search. The search for $r^*$ can be simply completed via exhaustive search method. The later numerical results in section \ref{s5} indicate that the total complexity of the above procedures is much less than that in \cite{zhu2013artificial}.

\subsection{Full ECSI at Alice}
When Alice gets an access to full ECSI, the SRM design is reformulated as
\begin{equation}\label{Cs2}
\begin{split}
&R_s^* = \max \;\ln \left| {{\mathbf{I}} + {{{\mathbf{\tilde H}}}_{ba}}{{\mathbf{Q}}_a}{\mathbf{\tilde H}}_{ba}^H} \right| - \ln \left| {{\mathbf{I}} + {{\mathbf{H}}_{{\rm{e}}a}}{{\mathbf{Q}}_a}{{\mathbf{H}}_{ea}}^H} \right|\\
&s.t.\quad {\rm{Tr}}({{\mathbf{Q}}_a}) \le P
\end{split}
\end{equation}
To solve (\ref{Cs2}), the AO algorithm proposed by \emph{Qiang Li} in \cite{li2013transmit} is utilized in our paper. Alternatively, one can adopt the GSVD algorithm proposed in \cite{fakoorian2012optimal} to deal with (\ref{Cs2}). The results of AO algorithm and GSVD algorithm will be presented in section \Rmnum{5}.
\begin{algorithm}
  \caption{The optimal strategy for Alice with statistical ECSI}
  \begin{algorithmic}[1]
    \State Initialize the number of Eve's antennas $N_E$, the total power budget $P$, the simulation number $N_{sim}$, the number of nonzero singular value of ${{\mathbf{\tilde H}}_{ba}}$ $c$.
    \State Generate $N_{sim}$ simulated channel matrices according to the distribution of $\mathbf{H}_{ea}$.
    \State ${{\rho }^{*}}\text{=}\arg \underset{\rho \in [0,1]}{\mathop{\max }}\,\left( \sum\limits_{i=1}^{{{N}_{sim}}}{\underset{r\in {{\mathbb{Z}}_{+}},1\le r\le c}{\mathop{\max }}\,{{{\hat{R}}}_{s}}(r,\rho ,i)} \right)$.
    \State ${{r}^{*}}\text{=}\arg \underset{r\in {{\mathbb{Z}}_{+}},1\le r\le c}{\mathop{\max }}\,\left( \sum\limits_{i=1}^{{{N}_{sim}}}{{{{\hat{R}}}_{s}}(r,{{\rho }^{*}},i)} \right)$.
    \State Compute the optimal power allocation matrix \textbf{p}, ${{\mathbf{p}}^{*}}\text{=waterfilling(}{{\rho }^{*}},{{r}^{*}}\text{)}$
    \State Choose $\mathbf{W}_{u}$ as the right singular vectors corresponding to the ${{r}^{*}}$ largest singular values of ${{\mathbf{\tilde{H}}}_{ba}}$. Choose $\mathbf{W}_z$ as the remaining ${{N}_{A}}-{{r}^{*}}$ right singular vectors of ${{\mathbf{\tilde{H}}}_{ba}}$
  \end{algorithmic}
\end{algorithm}

\section{Numerical Results}\label{s5}
In this section, we illustrate some numerical results to evaluate the performance gains of our proposed approaches. The parameters are set as: ${{N}_{A}}=6$, ${{N}_{E}}={{N}_{J}}={{N}_{B}}=4$, the total power budget $P=10\ {\rm{dB}}$. All channels follow an i.i.d. complex Gaussian distribution with zero mean and unit variance.

\begin{figure}[!t]
\centering
\includegraphics[width=8cm]{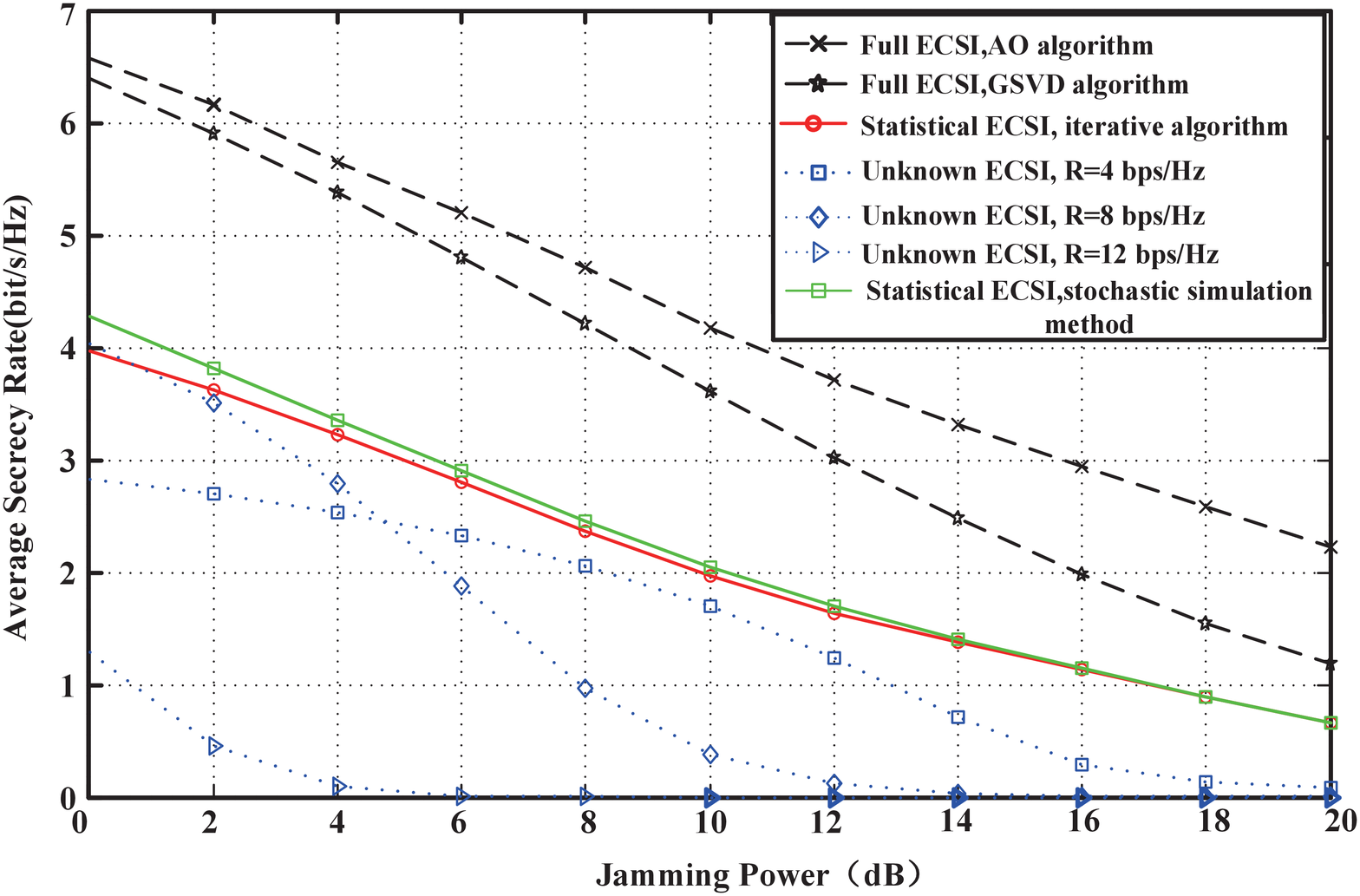}
\DeclareGraphicsExtensions.
\caption{Average secrecy rate versus the jamming power in various cases}\label{fig_all}
\end{figure}

\begin{figure}[!t]
\centering
\includegraphics[width=8cm]{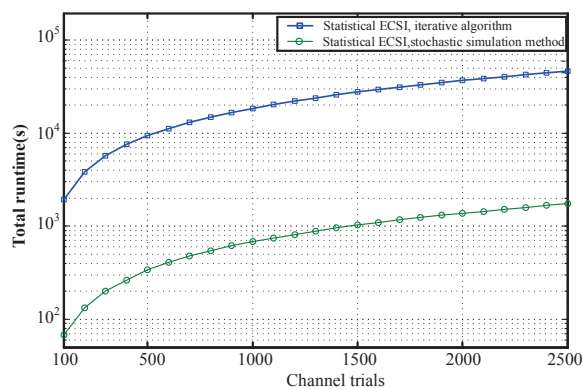}
\DeclareGraphicsExtensions.
\caption{Total runtime comparison between two methods in the case of statistical ECSI}\label{fig_complexity}
\end{figure}

Fig.\,\ref{fig_all} shows the performance of various algorithms designed for different levels of ECSI. We can see that better performance is always achievable with more ECSI. Particularly, the AO algorithm achieves the best performance among these algorithms. For the case of unknown ECSI, as seen, the average secrecy rate decreases significantly with the increment of jamming power. In addition, a lower target rate, $R=4$ bps/Hz, performs better than a higher target rate $R=8$ bps/Hz in high jamming power region. Namely, higher target rate means higher outage probability, thus leading to worse performance. As $R$ increases to 12 bps/Hz, the outage happens when the jamming power is only $6\ {\rm{dB}}$. For the case of statistical ECSI, we compare the performance of our stochastic simulation method with the iterative method proposed in \cite{zhu2013artificial}. We can find that our stochastic method even outperforms the iterative method in low jamming power region, also performs identically to the iterative method in high jamming power region. As for the reason, the iterative method only aims at maximizing the lower bound of the ergodic secrecy capacity while the stochastic method aims at maximizing an approximate ergodic secrecy capacity. Though the stochastic method requires myriads of simulated channels, the complexity is still much less than the iterative method, as shown in Fig.\,\ref{fig_complexity}.

In Fig.\,\ref{fig_complexity}, the runtime of stochastic search method can be $14\ {\rm{dB}}$ less than that of iterative algorithm in average. The notable runtime gap demonstrates the computational efficiency of stochastic search method. However, one can foresee that the stochastic method may be impractical when the scale of system keeps expanding, such as, in the massive MIMO system.

\section{Conclusion}
In this paper, we have studied a four-node wiretap channel with a malicious jammer. The system model is transformed into the transmitter-receiver-eavesdropper wiretap channel by whitening the interference at the legitimate receiver. With various levels of ECSI available to the transmitter, that is, full ECSI, unknown ECSI, and statistical ECSI, different transmission strategies are put forward to improve the secrecy performance. When no ECSI is available, we set a target rate for the legitimate receiver, and determine the power allocation matrix for information symbols by an inverse water-filling algorithm. A high target rate may encounter high outage probability, so the setting of target rate has to be prudent. When statistical ECSI is available, a stochastic simulation method is adopted to avoid the high complexity of iterative algorithm proposed in \cite{zhu2013artificial}. It can be seen that the stochastic simulation method can yield the performance identical to the iterative algorithm along with higher computational efficiency. Finally, when full ECSI is available, AO algorithm and GSVD algorithm, designed for conventional wiretap channel, are modified to solve the SRM problem in our paper.

\bibliography{vtc2015spring}
\bibliographystyle{IEEEtran}

\end{document}